\documentclass[
aps,%
12pt,%
final,%
notitlepage,%
oneside,%
onecolumn,%
nobibnotes,%
nofootinbib,%
superscriptaddress,%
noshowpacs,%
centertags]%
{revtex4}

\usepackage{epsf}

\usepackage{amsmath}
\usepackage{graphicx}
\usepackage[cp866]{inputenc}
\usepackage[english,russian]{babel}

\newcommand{\lan}{\langle}
\newcommand{\ran}{\rangle}

\newcommand{\be}{\begin{equation}}
\newcommand{\ee}{\end{equation}}
\newcommand{\ba}{\begin{aligned}}
\newcommand{\ea}{\end{aligned}}

\newcommand{\bea}{\begin{equation}\begin{aligned}}
\newcommand{\eea}{\end{aligned}\end{equation}}

\newcommand{\beq}{\begin{eqnarray}}
 \newcommand{\eeq}{\end{eqnarray}}

\def\fun#1#2{\lower3.6pt\vbox{\baselineskip0pt\lineskip.9pt
\ialign{$\mathsurround=0pt#1\hfil##\hfil$\crcr#2\crcr\sim\crcr}}}

\newcommand{{\PBC}}{{\rm PBC}}

\begin{document}


\title{Low-energy theorems of QCD and bulk viscosity at finite temperature and baryon density in a magnetic field \\ }
\author{N.O. Agasian}
\email{agasian@itep.ru}
\affiliation{
Institute of Theoretical and Experimental Physics, \\
117218, Moscow, Russia}


\begin{abstract}
The nonperturbative QCD vacuum at finite temperature and a finite baryon density in an external magnetic field is
studied. Equations relating nonperturbative condensates to the thermodynamic pressure for $T\neq 0$, $\mu_q \neq 0$ and $H\neq 0$
are obtained, and low-energy theorems are derived. A bulk viscosity $\zeta(T, \mu, H)$ is expressed in terms of
basic thermodynamical quantities describing the quark-gluon matter at $T\neq 0$, $\mu_q \neq 0$, and $H\neq 0$.
Various limiting cases are also considered.

\end{abstract}

\maketitle



\section{Introduction}

Quantum chromodynamics (QCD) as a theory that describes
strong interaction physics is still vigorously
developing. Over the past few decades, particular attention has
been drawn to the investigation of the behavior of strongly interacting
matter under the effect of various external fields. In the real
world, these are primarily temperature and the baryon density.
Interest in the behavior of matter under extreme conditions (high
temperature is the one starting from a characteristic QCD scale, $ T\sim
200 $ MeV,  and high baryon densities starts from $n> n_0 \simeq 0.17 $
fm$^{-3}$, where $n_0$ -- is a normal nuclear density) is
motivated primarily by the experiments
studying heavy-ion collisions. Due to this, one can expect that
such experiments probe densities and temperature at which a phase
transition to quark--gluon plasma, which is a new state of strongly
interacting matter, is possible.

In recent years, the phase structure of the vacuum in an external
magnetic field $H$ has become yet another important object of
investigations. It was shown recently that magnetic fields of
strength in the range $eH\sim 10 \div 10^4$ MeV$^2$ can be
generated in heavy-ion collisions \cite{Kharzeev:2007jp, Skokov:2009qp, Voronyuk:2011jd}.
Such field can lead to observable phenomena (so-called ``chiral magnetic effect'')
\cite{Kharzeev:1998kz, Kharzeev:2004ey, Kharzeev:2007jp,
Fukushima:2008xe, Fukushima:2010fe} in experiments at the
Relativistic Heavy Ion Collider (RHIC) and at the Large Hadron
Collider (LHC). Magnetic fields of the order of
$eH\sim \Lambda_{QCD}^2$ or even stronger could exist in the early Universe.
Such magnetic field strengths can lead to new interesting
phenomena accompanying the QCD  phase transition
\cite{Agasian:2008tb, Fraga:2008qn, Fraga:2008um, Mizher:2008hf, Buividovich:2008wf,
Ayala:2009ji, Buividovich:2009wi, Mizher:2010zb,
D'Elia:2010nq, Gatto:2010qs, Chernodub:2010qx, Ruggieri:2010zz,
Fayazbakhsh:2010bh, Gatto:2010pt, Ruggieri:2011qy, Frasca:2011zn,
D'Elia:2011zu, Andersen:2011ip, Andersen:2012dz, Andersen:2012bq,
Bali:2011qj, Braguta:2010ej, Buividovich:2010tn, Gorsky:2010xu, Kerbikov:2011gm}.

In \cite{Klevansky:1989vi}, the magnetic-field dependence of the
quark condensate was studied on the basis of the
Nambu--Jona-Lasinio model. In QCD, the one-loop result for the $H$
dependence of $\langle \bar q q\rangle$ was obtained in
\cite{Shushpanov:1997sf}. In both cases, the condensate was found
to grow with increasing $H$. It follows that a naive analogy with
superconductivity theory, where a magnetic field breaks the
condensate of Cooper pairs, is inapplicable here. Strictly
speaking, the behavior of the gluon condensate $\langle G^2\rangle$
in an Abelian magnetic field is also nontrivial. Gluons do not
carry an electric charge, but virtual quarks generated by them
shift the quantity $\langle G^2\rangle$ owing to their interaction with the
magnetic field $H$.
This phenomenon was studied in \cite{Agasian:1999vh} making use of the low-energy
theorems. The phase structure of QCD vacuum
in an Abelian magnetic field at finite temperature was
studied in \cite{Agasian:2000hw, Agasian:2008tb, Mizher:2010zb,
Gatto:2010qs, Ruggieri:2010zz, Bali:2011qj, Braguta:2010ej, Zayakin:2008cy}.

 Relations arising from the symmetry
properties play an important role in quantum field theories.
Searches for symmetries and constraints that these symmetries
impose on the physical properties of the system
become particularly important in QCD as a theory
that involves confinement, where composite states
(hadrons) appear to be ``observables''.
Low-energy theorems, or Ward identities (scale and chiral), play
a fundamental role in obtaining deeper insight into
the nonperturbative vacuum properties of QCD.
In QCD, low-energy theorems were obtained in
the early 1980s \cite{Novikov}.
The low-energy theorems of QCD, which follow from general symmetry properties
and which are independent of the details of the
confinement mechanism, make it possible to obtain
information that is sometimes inaccessible within any
other method.
Also, they can be used as ``physically reasonable'' constraints in constructing effective theories
and various models of the QCD vacuum.
In QCD, the low-energy theorems for $T\neq0$ and $\mu_q \neq 0$ were obtained in \cite{Ellis:1998kj, Shushpanov:1998ce}.
An important application of low-energy theorems in a hot QCD was obtained in \cite{Kharzeev:2007wb,
Karsch:2007jc}. Based on Kubo equation and low-energy theorems, the bulk viscosity of the quark-gluon matter  was
shown to be directly connected to a bilocal correlator of the energy-momentum tensor, and its value for a hot QCD
case was evaluated \cite{Kharzeev:2007wb, Karsch:2007jc}.

A method that makes it
possible to generalize the low-energy theorems of
QCD to the case of finite temperature, a finite baryon
chemical potential, and a nonzero magnetic field is
developed in the present study. This method is used
to study the nonperturbative vacuum and to derive
an expression for a bilocal correlation function for
the energy-momentum tensor and bulk viscosity in QCD for $T\neq0$, $\mu_q \neq 0$ and $H\neq0$.

\section{Low-energy theorems of QCD at finite $T, \mu$ and $H$}

In the Euclidean formulation, the QCD partition
function in presence of an external Abelian
field $A_\mu$ can be represented in the form (here, $T=1/\beta$ is temperature)
 \be
 Z=\exp \left \{ -\frac{1}{4e^2}
\int^\beta_0 dx_4\int_V d^3x F^2_{\mu\nu} \right \} \int[DB][D\bar q][Dq]
\exp \left \{ -\int^\beta_0 dx_4\int_V d^3x {\cal L} \right \},
 \label{eq_1}
  \ee
where the QCD Lagrangian in a background field has
the form
 \be
 {\cal L}=\frac{1}{4g^2_0}
 (G^a_{\mu\nu})^2+ \sum_{q=u,d,\ldots} \bar q[\gamma_\mu
 (\partial_\mu-iQ_q A_\mu-i\frac{\lambda^a}{2} B^a_\mu)+m_{0q}+\mu_q \gamma_0]q.
 \label{eq_2}
  \ee
Here, $Q_q$ is the charge matrix for quarks of flavor
 $q=(u,d,s,\ldots)$ and bare mass $m_{0q}$ and $\mu_q$ is the
quark chemical potential; the ghost and gauge-fixing
terms are not written down explicitly in order to avoid
encumbering the presentation.
The pressure in the system (minus a thermodynamical potential) is determined
by the expression $ \beta VP_0$ $(T,\mu_q, H, m_{0q})=\ln Z$.
From the partition function in Eq. (\ref{eq_1}), one can
obtain the following expression for the gluon condensate
 $\langle G^2\rangle\equiv \langle (G^a_{\mu\nu})^2\rangle$:
 \be
 \langle G^2\rangle (T,\mu_q, H,m_{0q})=-4\frac{\partial P_0}{\partial(1/g^2_0)}~.
 \label{eq_4}
 \ee
The system described by the partition function in (\ref{eq_1})
is characterized by the set of dimensional parameters $\Lambda_0, \mu_q, T, H, m_{0q}(\Lambda_0)$
and by the dimensionless charge  $g^2_0(\Lambda_0)$, where the bare quark masses $m_{0q}$
and the coupling constant $g^2_0$ are specified at the
scale of the ultraviolet-cutoff mass $\Lambda_0$.
On the other hand, we can go over to the renormalized (physical)
pressure $P$ and, with the aid of the properties
of the renormalization group invariance of $P$,
recast expression (\ref{eq_4}) into a form that involves derivatives
with respect to the physical parameters
$T, \mu_q,$ and  $H$ and with respect to the renormalized masses $m_q$.

The dimensional transmutation phenomenon leads
to the appearance of the nonperturbative dimensional
parameter
  \be
  \Lambda= \Lambda_0 \exp
 \left \{ \int^\infty_{\alpha_s(\Lambda_0)}\frac{d\alpha_s}{\beta(\alpha_s)}
 \right \},
  \label{eq_5}
  \ee
where $\Lambda_0$ is the ultraviolet-cutoff mass, $\alpha_s=g^2_0/4\pi$ and $\beta(\alpha_s)=d\alpha_s(\Lambda_0)/d\ln \Lambda_0$
is theGell-Mann-Low function.
The quark mass $m_{0q}$ has the anomalous dimension $\gamma_{m_q}$ and depends on the scale $\Lambda_0$.
The renormalization group equation for the running mass  $m_{0q}(\Lambda_0)$ has the form
 $d\ln m_{0q}/d\ln \Lambda_0=-\gamma_{m_q}$, and we use the modified minimal-subtraction $\overline{MS} $
scheme, where $\beta$ and $\gamma_{m_q}$ are independent of the quark mass.
The expression for the renormalization group invariant mass then has the form
\be
m_q=m_{oq}(\Lambda_0)\exp \left \{\int^{\alpha_s(\Lambda_0)}\frac{\gamma_{m_q}(\alpha_s)}{\beta(\alpha_s)}
d\alpha_s \right \}.
 \label{eq_6}
 \ee
Since the physical (renormalized) pressure is a renormalization group invariant quantity, its anomalous
dimension is zero. Thus,  $P$ has only a normal (canonical) dimension equal to four.
Employing the renormalization group invariance of the quantity $\Lambda$, we can write $P$
in the most general form as
\be P=\Lambda^4 f(\frac{T}{\Lambda}, \frac{\mu_q}{\Lambda},
\frac{H}{\Lambda^2}, \frac{m_q}{\Lambda}),
 \label{eq_7}
 \ee
where  $f$ is a function of the dimensionless ratios   $T/\Lambda,...$.
From (\ref{eq_5}),(\ref{eq_6}) and (\ref{eq_7}) we then obtain
\be
\frac{\partial P}{\partial(1/g^2_0)} =\frac{\partial
P}{\partial\Lambda} \frac{\partial\Lambda}{\partial(1/g^2_0)} +
\sum_q \frac{\partial P}{\partial m_q} \frac{\partial
m_q}{\partial(1/g^2_0)}~,
 \label{eq_8}
 \ee

\be
\frac{\partial m_q}{\partial(1/g^2_0)}=-4\pi\alpha^2_s
m_q\frac{\gamma_{m_q}(\alpha_s)}{\beta(\alpha_s)}~.
 \label{eq_9}
 \ee
Further, The anomaly in the trace of the energy-momentum tensor in QCD is related to the gluon
condensate by the equation
\be
\lan \theta_{\mu \mu}^g \ran = \frac{\beta(\alpha_s)}{16\pi\alpha_s^2}\lan G^2\ran.
\label{eq_9_a}
 \ee
Taking into account (\ref{eq_4}), we obtain the gluon part of the trace of the energy-momentum tensor in the form
\be
\lan \theta_{\mu \mu}^g \ran
=\left(T\frac{\partial}{\partial T}+\sum_q {\mu_q} \frac{\partial}{\partial {\mu_q}} +2 H\frac{\partial}{\partial
H}+\sum_q(1+\gamma_{m_q})m_q\frac{\partial}{\partial {m_q}}-4 \right) P.
\label{eq_10_0}
 \ee
 Here and below, the energy-momentum tensor $\lan \theta_{\mu \mu} \ran$,
the condensates $\langle G^2\rangle$ and $\lan\bar q q\ran$,
and the thermodynamic pressure $P$ are functions of $T, \mu_q, H$ and $m_q$.

In the one-loop approximation, we have  $\beta(\alpha_s)\to - b\alpha^2_s/2\pi$ and $1+\gamma_{m_q}\to 1$, where $b=(11 N_c-2N_f)/3$.
Thus, the gluon and quark parts of the trace of the energy-momentum tensor in hot
and dense QCD in a magnetic field can be expressed in terms of the physical pressure in the one-loop approximation as
\be
\lan \theta_{\mu \mu}^g \ran = - \frac{b}{32\pi^2} \lan G^2\ran =
\left(T\frac{\partial}{\partial T}+\sum_q {\mu_q} \frac{\partial}{\partial {\mu_q}}
+2H\frac{\partial}{\partial H}+\sum_q m_q\frac{\partial}{\partial m_q}-4 \right) P,
 \label{eq_11_0}
 \ee

 \be
\lan \theta_{\mu \mu}^q \ran = \sum_q m_q \lan\bar q q\ran =-\sum_q m_q \frac{\partial P}{\partial {m_q}}~.
 \label{12_0}
 \ee

In the vacuum, that is, at $T=0$, $\mu_q=0$ and $H=0$, we arrive at the well-known expression for
the nonperturbative vacuum energy density. In the one-loop approximation, this expression has the form
$$
\varepsilon_{\rm vac}=\frac{1}{4}\lan {\theta_{\mu \mu}^g + \theta_{\mu \mu}^q }\ran_0=-P(T=0,\mu_q=0,H=0, m_q)
$$
\be
= -\frac{b}{12 8 \pi^2} \langle G^2\rangle_0 + \frac{1}{4} \sum_q m_q \lan\bar q q\ran_0.
\label{9_gl4}
\ee

By using the relations presented above, one can
derive low-energy theorems of QCD at finite temperature
and finite density in the presence of a magnetic
field. Strictly speaking, the $\beta$-function  depends
on $H$, so that the low-energy theorems could involve
electromagnetic corrections, which are proportional
to $\propto e^4$, but, since the physical pressure is independent
of the scale $\Lambda_0$ at which ultraviolet divergencies are
regularized, one can choose an ultraviolet scale in
such a way that $\Lambda_0\gg H, T, \mu_q, \Lambda$.
We can then restrict ourselves to the lowest order in the expansion of
the $\beta$-function, with the result that electromagnetic
corrections disappear. Taking this into account, we
consider the trace of the energy-momentum tensor
in hot and dense QCD in the one-loop approximation,
\be
\theta_{\mu \mu} (x) = - \frac{b}{32 \pi^2} (G^a_{\mu\nu}
(x))^2+\sum_q m_q \bar q q (x).
\label{eq_10.a}
\ee
Also, we introduce the operator $\hat D$, defining it as
\be \hat D=T\frac{\partial}{\partial T}
+\sum_q {\mu_q} \frac{\partial}{\partial {\mu_q}}+2H\frac{\partial}{\partial H}.
\label{eq_11.a}
 \ee
From relations (\ref{eq_11_0}) and (\ref{12_0}), we find for the total
vacuum expectation value of the trace of the energy-momentum tensor with allowance for massive quarks that
\be
\lan \theta_{\mu \mu} \ran = \lan \theta_{\mu \mu}^g  +\theta_{\mu \mu}^q \ran= - \frac{b}{32\pi^2} \lan G^2\ran
+\sum_q m_q \lan\bar q q\ran =(\hat D -4)P.
 \label{eq_13_b}
 \ee
Differentiating Eq.(\ref{eq_4}) $n$ times with respect to $(1/g^2_0)$ and taking into account relations
(\ref{eq_7}), (\ref{eq_10.a}), (\ref{eq_11.a}) and (\ref{eq_13_b}), we obtain
$$
(\hat D-4)^{n+1} P=(\hat D-4)^n\langle \theta_{\mu \mu}^g (0)\rangle
$$
\be
=\int d^4 x_n... \int d^4 x_1 \langle
\theta_{\mu \mu}^g (x_n)...
\theta_{\mu \mu}^g (x_1)
\theta_{\mu \mu}^g (0)\rangle.
\label{eq_12.a}
\ee
To the right-hand side of (\ref{eq_12.a}) only connected diagrams are included, as usual.

 Similar arguments apply to an arbitrary
operator $\hat{O}(x)$ constructed from quark or gluon fields; that is,
 $$
 \left(T\frac{\partial}{\partial T}+\sum_q {\mu_q} \frac{\partial}{\partial {\mu_q}} + 2 H
 \frac{\partial}{\partial H} -d \right)^n\langle \hat{O}\rangle
 $$
 \be
 =\int d^4 x_n...
 \int d^4 x_1
\langle \theta_{\mu \mu}^g (x_n)...
 \theta_{\mu \mu}^g (x_1)
 \hat{O}(0)\rangle,
 \label{eq_13.a}
 \ee
 where $d$ is the canonical dimension of the operator $\hat{O}$. If the operator $\hat{O}$
 has an anomalous dimension as well, it is necessary to take into account the corresponding
$\gamma$-function.

Let us now consider the case of $n=1$, which is of
importance for physical applications. In other words,
we will examine a bilocal correlation function for
the tensors of the energy-momentum density in hot
and dense QCD in a magnetic field. In terms of this
correlation function, one can express the bulk viscosity of
quark--gluon plasma in a magnetic field.
Then, for gluon and quark contributions to the bilocal correlator of the trace of the energy-momentum tensor, we have
the following relations:
\be
 \int d^4 x \lan \theta^g_{\mu\mu} (x) \theta^g_{\mu\mu}
(0)\ran = (\hat D-4) \lan \theta^g_{\mu\mu}\ran,
\label{17a}
\ee
\be
\int d^4 x \lan \theta^g_{\mu\mu} (x) \theta^q_{\mu\mu}
(0)\ran = (\hat D-3) \lan \theta^q_{\mu\mu}\ran.
\label{17}
\ee
Hence, for the the bilocal correlator of the trace of the energy-momentum tensor
$$
\Pi = \int d^4 x \lan \theta_{\mu\mu} (x) \theta_{\mu\mu}(0)\ran
$$
\be
 =\int d^4 x \lan \theta^g_{\mu\mu}
\theta^g_{\mu\mu}\ran + 2 \int d^4 x \lan \theta^g_{\mu\mu}
\theta^q_{\mu\mu}\ran  + O(m^2_q),
\label{18}
\ee
where we included a correlator of quark summands to $O(m^2_q)$, being proportional to a quark mass
squared; in what follows we will not take it into account. Based on relations (\ref{17}), we find
\be
\Pi= (\hat D-4) \lan \theta^g_{\mu\mu}\ran+2(\hat D-3) \lan
\theta^q_{\mu\mu}\ran
=(\hat D-4) \lan \theta_{\mu\mu}\ran +(\hat D-2) \lan
\theta^q_{\mu\mu}\ran.
\label{19}
\ee

\section{Bulk viscosity $\zeta(T, \mu, H)$}

As it was shown in \cite{Karsch:2007jc}, according to a general Kubo formula, a bulk viscosity can be
evaluated as a static limit of a bilocal correlator of the trace of the energy-momentum tensor.
\be
\zeta = \frac{1}{9}\lim_{\omega\to 0}\frac{1}{\omega}\int_0^\infty dt \int d^3r\,e^{i\omega t}\,
\langle [\theta_{\mu\mu}(x),\theta_{\mu\mu}(0)]\rangle \,.
\label{kubo}
\ee
One can introduce a spectral density expressed in terms of a retarded Green function for the
trace of the energy-momentum tensor
\be
\rho(\omega,\mathbf{p})=-\frac{1}{\pi}\mbox{Im} G^R(\omega,\mathbf{p}).
\label{spectral1}
\ee
Then, as was suggested in \cite{Karsch:2007jc}, for small frequencies the spectral density has the
following form:
\be
\frac{\rho(\omega,\mathbf{0})}{\omega}=\frac{9\zeta}{\pi} \frac{\omega_0^2}{\omega_0^2+\omega^2},
\label{spectral2}
\ee
where the $\omega_0$ parameter determines a scale at which a perturbation theory becomes applicable.
Using this ansatz, the bulk viscosity can be written as
\be
9\zeta\omega_0=2\int \limits_0^\infty \frac{\rho(u,\mathbf{0})}{u}\,du
=\int d^4 x \lan \theta_{\mu\mu} (x) \theta_{\mu\mu}(0)\ran=\Pi.
\label{v1}
\ee
Thus, a problem to find the bulk viscosity $\zeta$ reduces to a problem to evaluate the bilocal
correlator $\Pi$.

We extract from the correlator $\Pi$ a vacuum term. For this purpose, we write the following expression
for the total pressure
\be
P=-\varepsilon_{\rm vac} + P_*,
\label{20}
\ee
where $P_*$ is the pressure pure thermodynamical part. Quark and gluon contributions to the
trace of the energy-momentum tensor can be written as
$$
\lan \theta^q_{\mu\mu}\ran = \lan \theta^q_{\mu\mu}\ran_0+ \lan
\theta^q_{\mu\mu}\ran_* = \sum_q m_q \lan \bar q q\ran_0 + \sum_q m_q \lan \bar
q q\ran_*
$$
\be
\lan \theta^g_{\mu\mu}\ran= \lan \theta^g_{\mu\mu}\ran_0+ \lan
\theta^g_{\mu\mu}\ran_*
\label{21}
\ee
and, using equation (\ref{eq_13_b})
\be
\lan \theta_{\mu\mu}\ran= \lan \theta^q_{\mu\mu}+ \theta^g_{\mu\mu}\ran=4\varepsilon_{\rm vac} + (\hat  D-4)
P_*
\label{22}
\ee
Allowing for the thermodynamic relation
\be
\left(T\frac{\partial}{\partial T} + \sum_q  \mu_q \frac{\partial}{\partial \mu_q}-4\right) P=\varepsilon- 3P
\label{23}
\ee
and taking into consideration the magnetic moment $M= {\partial P}/{\partial H}$
we have
\be
\lan \theta_{\mu\mu}\ran =4\varepsilon_{\rm vac}+(\varepsilon- 3 P)_*+2MH.
\label{24}
\ee
Then the correlator (\ref{18}, \ref{19}) can be written as
\be
\Pi =\Pi_0 + \Pi^q_* + \Pi_*^g,
\label{25}
\ee
where the vacuum contribution
\be
\Pi_0 =-4\lan \theta_{\mu\mu}\ran_0 - 2 \lan \theta^q_{\mu\mu}\ran_0 = -16 \varepsilon_{\rm vac} - 2\sum_q
m_q \lan \bar q q\ran_0.
\label{26}
\ee
For the quark contribution $\Pi^q_*$ we get the following expression
\be
\Pi^q_* =\left(T\frac{\partial}{\partial T}+ \sum_q \mu_q \frac{\partial}{\partial \mu_q}+ 2H
\frac{\partial}{\partial H}-2\right)\sum_q m_q \lan \bar q q \ran_*
\label{29}
\ee
The gluonic part $\Pi^g_*$ of the correlator can be written as
$$
\Pi^g_*= \left(T\frac{\partial}{\partial T}-4\right) (\varepsilon - 3 P)_*
+\left(\sum_q  \mu_q \frac{\partial}{\partial \mu_q}+ 2H
\frac{\partial}{\partial H}\right)(\varepsilon - 3 P)_*
$$
\be
+\left(T\frac{\partial}{\partial T}+ \sum_q  \mu_q \frac{\partial}{\partial \mu_q}+
2H \frac{\partial}{\partial H}-4\right) 2MH.
\label{30}
\ee
We use the following definitions of thermodynamic quantities, in terms of pressure and energy density:
entropy density $s={\partial P}/{\partial T}$, specific heat  $c_v={\partial \varepsilon}/{\partial T}$,
velocity of sound $c^2_s={\partial P}/{\partial \varepsilon}=s/c^2_v$, and medium magnetic susceptibility $\chi={\partial
M}/{\partial H}={\partial^2 P}/{\partial H^2}$. Then we find
$$
\Pi^g_* = T s \left(\frac{1}{c^2_s} -3\right) +\left(\sum_q  \mu_q
\frac{\partial}{\partial \mu_q}-4\right)(\varepsilon - 3 P)_*
$$
\be
+ 4 \chi H^2 -12 MH +
4H\left(T\frac{\partial}{\partial T}+\sum_q  \mu_q \frac{\partial}{\partial
\mu_q}\right) M.
\label{31}
\ee
Thus, equations (\ref{26}),  (\ref{29}), and (\ref{31}) express the bilocal correlator of the trace of the
energy-momentum tensor $\Pi$, and, correspondingly, in accordance with (\ref{v1}), of the bulk viscosity
$\zeta$ via thermodynamic parameters: $T$, $\mu_q$, $H$, $\lan \bar q q \ran_*$, $(\varepsilon - 3 P)$, $s$, $c^2_s$, $M$, $\chi$

$$
9\zeta(T, \mu, H)\omega_0 = -16\varepsilon_{\rm vac} - 2\sum_q m_q \lan \bar q q\ran_0
$$
$$
+\left(T\frac{\partial}{\partial T}+
\sum_q  \mu_q \frac{\partial}{\partial \mu_q}+ 2H \frac{\partial}{\partial
H}-2\right)\sum_q m_q \lan \bar q q \ran_*
$$
$$
+ T s \left(\frac{1}{c^2_s} -3\right) +\left(\sum_q  \mu_q
\frac{\partial}{\partial \mu_q}-4\right)(\varepsilon - 3 P)_*
$$
\be
+ 4 \chi H^2 -12 MH +
4H\left(T\frac{\partial}{\partial T}+\sum_q  \mu_q \frac{\partial}{\partial
\mu_q}\right) M.
\label{32}
\ee

We explore various limiting cases. Let us consider $\mu_q=H=0$ and $T\neq 0$, which corresponds to hot QCD
studied in \cite{Karsch:2007jc}. Taking into account (\ref{9_gl4}) and using PCAC relation
\be
\sum_q m_q \lan \bar q q\ran_0= - F^2_\pi M^2_\pi - F^2_KM^2_K
\label{33}
\ee
we write the vacuum contribution $\Pi_0$ as
\be
\Pi_0 =-4\lan \theta_{\mu\mu}\ran_0 - 2 \lan \theta^q_{\mu\mu}\ran_0
=-4\lan \theta^g_{\mu\mu}\ran_0 - 6\lan \theta^q_{\mu\mu}\ran_0
= 16 |\varepsilon^g_{\rm vac}| + 6 (F^2_\pi M^2_\pi +F^2_KM^2_K),
\label{35}
\ee
where $\varepsilon^g_{\rm vac}$ is the gluonic contribution to the vacuum energy density (\ref{9_gl4}).
Assigning  $\mu_q=H=0$ in  (\ref{32})  and allowing for (\ref{33}), (\ref{35}), we find for the
correlator $\Pi$ and , correspondingly, for the bulk viscosity
$$
9\zeta(T)\omega_0= T s \left(\frac{1}{c^2_s} -3\right)-4(\varepsilon - 3 P)_*
+\left(T\frac{\partial}{\partial T}-2\right)\sum_q m_q \lan \bar q q \ran_*
$$
\be
+16 |\varepsilon^g_{\rm vac}| + 6 (F^2_\pi M^2_\pi +F^2_KM^2_K),
\label{36}
\ee
which exactly conforms to the main result in \cite{Karsch:2007jc}.

Now consider the utmost case of the cold quark matter $T=H=0$ and $\mu_q\neq 0$. And take into account
that for such a variant
\be
\varepsilon- 3 P=\left( \sum_q  \mu_q \frac{\partial}{\partial \mu_q}-4\right) P =\sum_q \mu_q n_q -4P,
\label{37}
\ee
where $n_q={\partial P}/{\partial \mu_q}$ is the quark density. We use relation (\ref{21}) and definition
$P(\mu)=-\varepsilon_{\rm vac}+P_*$, then we suppose $T=H=0$ in (\ref{32}) , and, after simple transformations,
we find\footnote{The expression obtained for the bulk viscosity of the cold quark matter (\ref{38}) does not
coincide with a corresponding result in \cite{Wang:2011ur}.}

\be
9\zeta(\mu)\omega_0 = 16P(\mu)-2\sum_q m_q \lan \bar q q\ran+\sum_q m_q \mu_q \frac{\partial \lan \bar q
q\ran}{\partial \mu_q} -7 \sum_q \mu_q n_q + \sum_{q, q\prime}  \mu_q  \mu_{q\prime} \frac{\partial^2
P(\mu)}{{\partial \mu_q}{\partial \mu_{q\prime}}}
\label{38}
\ee
We note that in (\ref{38}) an expression for the all quark condensate $\lan \bar q q\ran$ including
vacuum term is entering.

In the limiting case of hot and dense quark-gluon matter, an expression for  $\zeta(T, \mu, H=0)$ is
obtained immediately from equation (\ref{32}) for $\zeta(T, \mu, H)$ by substituting $H=0$.

\section{Conclusions}

According to Eq. (\ref{32}), the bulk viscosity $\zeta$  in a
magnetic field is proportional to the magnetic susceptibility $\chi$.
In two different phases of strongly interacting matter (hadron and quark-gluon), the magnetic
susceptibility has different signs \cite{Agasian:2008tb}.
This is because hadron matter at temperatures below
the quark-hadron phase transition primarily consists of a gas of hot $\pi$-mesons,
which in the magnetic field behaves as a diamagnetic medium of spinless charged particles. The
system above the phase transition point is in the paramagnetic phase of hot quarks and gluons.
This significantly affects the bulk viscosity of strongly interacting
matter in the magnetic field, which should have a jump
at the transition point, leading to interesting observable phenomena.

In this paper a nonperturbative QCD vacuum at a finite temperature and a finite baryon chemical
potential in an external magnetic field are considered. Relations between non-perturbative condensates and
thermodynamical pressure at $T\neq 0$, $\mu_q \neq 0$, and $H\neq 0$ are obtained, and low-energy theorem are
derived. A common formula for the bulk viscosity of the quark-gluon medium at $T\neq 0$, $\mu_q \neq 0$,
and $H\neq 0$, is derived, which links $\zeta(T, \mu, H)$ to thermodynamic system parameters: entropy density
$s={\partial P}/{\partial T}$,  velocity of sound $c^2_s={\partial P}/{\partial \varepsilon}=s/c^2_v$,
non-ideality $(\varepsilon - 3 P)$, quark condensate $\lan \bar q q \ran_*$, and medium magnetic
susceptibility $\chi={\partial M}/{\partial H}={\partial^2 P}/{\partial H^2}$. Some physically interesting limiting cases
for the bulk viscosity are considered.
These phenomena and exact relations obtained here may prove to be of importance in examining the quark--hadron
phase transition in heavy ion collisions and in the early Universe, where strong magnetic fields are generated.

I am grateful to A.B. Kaidalov, V.A. Novikov, and Yu.A. Simonov for useful discussions about the problems
considered in the present study.


\end{document}